\documentclass[aps,prb,twocolumn,showpacs,preprintnumbers,amsmath,amssymb,superscriptaddress]{revtex4}

\usepackage{graphicx}
\usepackage{dcolumn}
\usepackage{bm}
\usepackage{figsize}

\begin{document}


\title{Effects of Spin Polarization in the HgTe Quantum Well}

\author{M. V. Yakunin}
\email[]{yakunin@imp.uran.ru}
\affiliation{Institute of Metal Physics, S. Kovalevskaya Str., 18, Ekaterinburg 620990, Russia}
\affiliation{Ural Federal University, Mira Str.,19, Ekaterinburg 620002, Russia}
\author{A. V. Suslov}
\affiliation{NHMFL, FSU, 1800 East Paul Dirac Drive, Tallahassee, Florida 32310, USA}
\author{S.~M.~Podgornykh}
\affiliation{Institute of Metal Physics, S. Kovalevskaya Str., 18, Ekaterinburg 620990, Russia}
\affiliation{Ural Federal University, Mira Str.,19, Ekaterinburg 620002, Russia}
\author{S. A. Dvoretsky}
\affiliation{Institute of Semiconductor Physics, Lavrentyev Ave., 13, Novosibirsk 630090, Russia}
\author{N. N. Mikhailov}
\affiliation{Institute of Semiconductor Physics, Lavrentyev Ave., 13, Novosibirsk 630090, Russia}

\date{\today}

\begin{abstract}
Magnetoresistivity features connected with the spin level coincidences under tilted fields  in a $\Gamma_8$ conduction band of the HgTe quantum well were found to align along straight trajectories in a $(B_\bot,B_{||})$ plane between the field components perpendicular and parallel to the layer meaning a linear spin polarization dependence on magnetic field. Among the trajectories is a noticeable set of lines descending  from a single point on the $B_{||}$ axis, which  is shown to yield a field of the full spin polarization of the electronic system, in agreement with the data on the electron redistribution between spin subbands obtained from Fourier transforms of oscillations along circle trajectories in the $(B_\bot,B_{||})$ plane and with the point on the magnetoresistivity under pure $B_{||}$ separating a complicated weak field dependence from the monotonous one. The whole picture of coincidences is well described by the isotropic $g$-factor although its value is twice as small as that obtained from oscillations under pure perpendicular fields. The discrepancy is attributed to different manifestations of spin polarization phenomena in the coincidences and within the exchange enhanced spin gaps. In the quantum Hall range of $B_\bot$, the spin polarization manifests in anticrossings of magnetic levels, which were found to depend dramatically nonmonotonously  on $B_\bot$.
\end{abstract}

\maketitle

\section{\label{sec:level1}Introduction}
Spin polarization underlies the basic principles of spintronics devices. In addition, in the samples of quantum Hall quality, it opens up a rich and varied phenomenology, giving us the opportunity to investigate the role of interparticle interactions, and may lead to the formation of ordered many-particle ground states, including ferromagnetic ones.\cite{Jungwirth-McD} Full spin polarization of an electronic system is easily achievable in the conduction band of  $\Gamma_8$ nature in the inverted energy band spectrum of the HgTe quantum well (QW) wider than 6.3~nm (Ref.~\onlinecite{Konig}) due to a large value of $g^*m^*/m_0$ (the effective Lande $g$-factor multiplied by the effective to free mass ratio) thus making this material  promising for observation of a variety of spin phenomena. In particular, rich patterns of the spin level coincidences are observed in HgTe QW under tilted magnetic fields that extend into high perpendicular fields where the quantum Hall effect (QHE) is well realized.\cite{YaPE,Ya-prev} The existence of the coincidence effect in the $\Gamma_8$ conduction band of the HgTe QW has been revealed in Ref.~\onlinecite{Zhang-2004} and, although the detailed picture of magnetic levels under tilted magnetic fields was unknown in this case, the field-independent effective $g$-factor has been found for a symmetrically doped (001) oriented structure when analyzing the results in terms of a $\Gamma_6$-like band. In this paper, we develop this study for the (013) oriented symmetrical HgTe QWs with lower electron densities and larger mobilities, extending them to higher perpendicular fields and to the lowest integer filling factors with the use of experimental technique for obtaining detailed pictures of magnetoresistivities (MRs) as functions of two variables -- the perpendicular and parallel field components: $\rho_{xx,xy}(B_{\bot},B_{||})$.\cite{YaJETP} This resulted in finding a number of peculiar behaviors in quantum magnetotransport under tilted magnetic fields: a new system in locations of the coincidence features on the $(B_{\bot},B_{||})$-plain that yields a field of full spin polarization; an unusual connection between oscillations under pure perpendicular and tilted fields; a peculiar behavior of MR at field orientations close to parallel to the layers; and the existence of anticrossings in the quantum Hall range of perpendicular fields, which display a phase-transition-like behavior with changes in the sample quality and dramatic nonmonotonous changes with the filling factor. These findings are useful to compare with similar results obtained recently on the InSb QW, which has a comparable value of $g^*m^*/m_0$ but in the $\Gamma_6$ conduction band.\cite{Nedniyom,Yang-Santos} 

\section{\label{sec:level2}Experimental results}

\begin{figure}[t]
\includegraphics[width=\columnwidth]{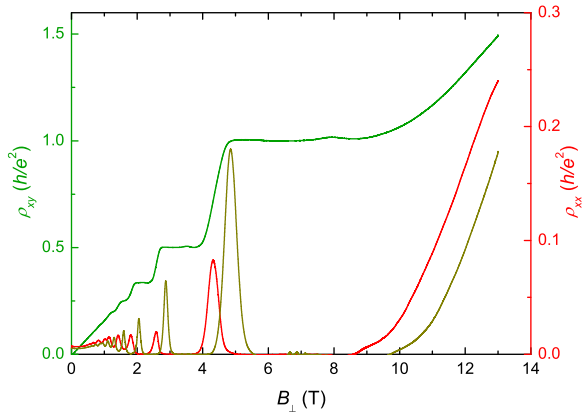}
\caption{\label{fig:BperXX-XY}  (color online). Quantum Hall effect under pure perpendicular field before and after ($\rho_{xx}$ shifted right) an extreme illumination. $T=0.32~K$. Notice a considerable increase of the peak amplitudes after illumination.}
\end{figure}

We present a study of quantum magnetotransport in a 20.3-nm-wide HgTe QW grown on the (013) GaAs substrate, symmetrically modulation doped with In at both sides at distances of about 10-nm spacers. The electron gas density is $n_S=1.5\times10^{15}$ m$^{-2}$ with a mobility of 22~m$^2$/Vs. The sample is in a shape of a Hall bar with Ohmic contacts. Longitudinal and Hall magnetoresistances were measured simultaneously by the direct current reversal technique using our original experimental setup \cite{Suslov} on a dc current of $1\div2$~$\mu$A at magnetic fields up to 18~T at temperatures of 0.32~K and higher. The sample was mounted within an {\it in situ} low friction rotator\cite{Palm} and the whole installation allowed measurements of MR as a continuous function of the tilt angle at the lowest temperatures so as not to cause a heating of the system. The structure reacts to the ir illumination with a persistent increase in $n_S$ up to about 20\%, but the main effect of illumination is a considerable improvement of the sample quality as the peak amplitudes increase more than twice while the zero-field resistivity drops down (Fig.~\ref{fig:BperXX-XY}). Measurements in tilted magnetic fields were organized in a way to span in detail the whole available circle area in the  $(B_{\bot},B_{||})$-plane to build the longitudinal and Hall MRs as continuous functions of two variables $\rho_{xx,xy}(B_{\bot},B_{||})$\cite{YaJETP}:  Figures~\ref{fig:XX-3D anticross},~\ref{fig:DescTraj},~\ref{fig:XX-XY}, \ref{fig:XX_g-anisotr}, \ref{fig:Bpar}, \ref{fig:IL1-CS1(nu3)cropped}.

\subsection{\label{sec:level3}Spin level coincidences as a tool to determine spin polarization}

\begin{figure}[b]
\includegraphics[width=9.5cm]{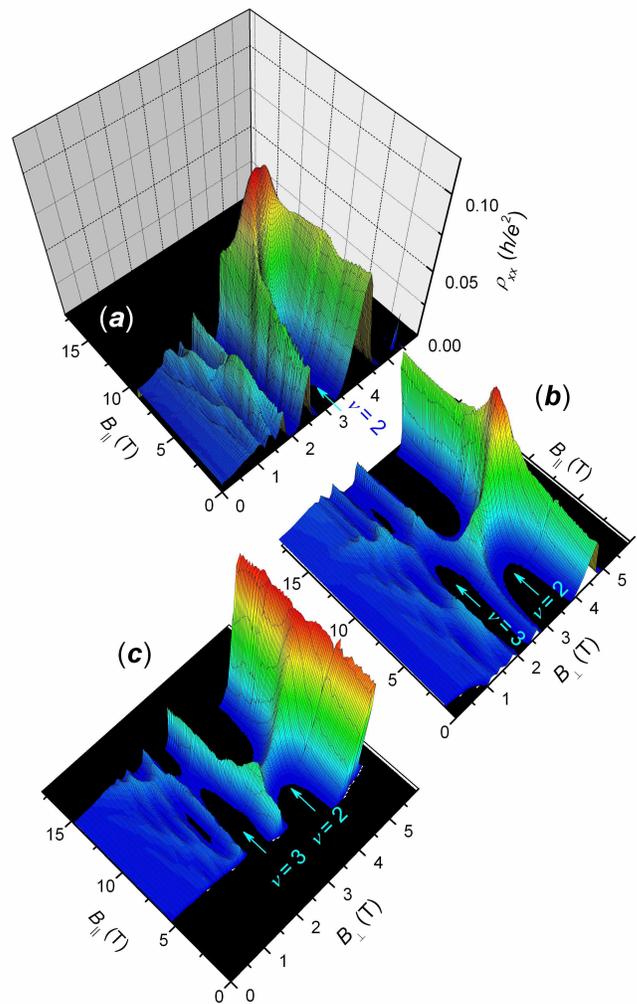}
\caption{\label{fig:XX-3D anticross}  (color online). Evolution of the coincidence features in $\rho_{xx}(B_{\bot},B_{||})$ with illumination at $T=0.32$~K: (a) before illumination; (b) after an intermediate and (c) extreme illuminations.}
\end{figure}

\begin{figure}[b]
\includegraphics[width=9.5cm]{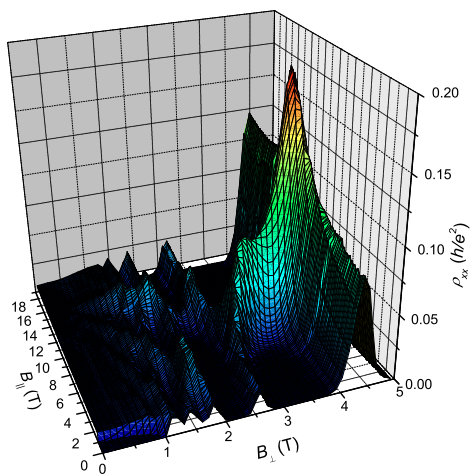}
\caption{\label{fig:DescTraj}  (color online). The same as in Fig.~\ref{fig:XX-3D anticross}(b), but at different angle of view, with the descending ridges formed by the coincidence peaks well visible.}
\end{figure}

\begin{figure}[b]
\includegraphics[width=9cm]{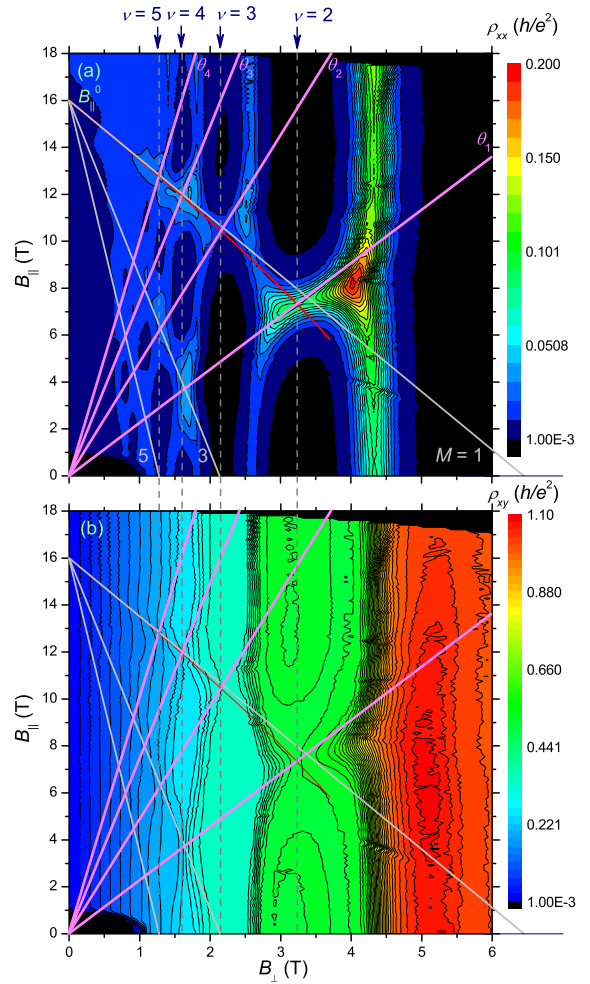}
\caption{\label{fig:XX-XY} (color online).  (a) The same as in Fig.~\ref{fig:XX-3D anticross}(b), but shown as a map, and (b) a corresponding $\rho_{xy}(B_{\bot},B_{||})$. The traditional trajectories at fixed angles $\theta_{r}$ and the second type (descending) trajectories for $M=1,3,5$ are drawn according to Eqs. (\ref{eq:trad}) and (\ref{eq:Bpar-Bper}). The line declining from the $M=1$ beam is a result of the exact calculation.}
\end{figure}

Spin level coincidences manifest themselves as sharp transformations of $\rho_{xx}(B_{\bot},B_{||})$ minima into local peaks (Figures~\ref{fig:XX-3D anticross},~\ref{fig:DescTraj},~\ref{fig:XX-XY}(a)) and concomitant local smoothings of the QH plateaux in $\rho_{xy}(B_{\bot},B_{||})$ (Fig.~\ref{fig:XX-XY}(b)) at the integer filling factors $\nu$. These coincidence features reside on trajectories of different kinds in the $(B_{\bot},B_{||})$-plane. First, they align with a good accuracy on a set of straight beams going up from zero with tilt angles $\theta_r$ satisfying a traditional relation for the coincidences\cite{Nicholas}: 
\begin{equation}
g^*m^*/m_0=2r\cos{\theta_r},
\label{eq:trad}
\end{equation}
with $r= 1,2,3\dots$ being the ratios of the spin to cyclotron gaps. This means that the coincidences in the $\Gamma_8$ conduction band in our case may be well described in terms of a simple $\Gamma_6$ band in a traditional semiconductor QW within a one-particle approach, when the cyclotron gaps depend only on $B_{\bot}$ and the spin gaps depend on the total field $B$, as in Ref.~\onlinecite{Zhang-2004}. This occurs in spite of the known complicated picture of magnetic levels of the $\Gamma_8$ conduction band under perpendicular fields\cite{Konig} and we treat the found similarity of its behavior under tilted fields with that for a  $\Gamma_6$ band as an empirical fact until it is explained by detailed calculations of the magnetic level structure under tilted fields. 

It is notable that the observed linear shape of these trajectories is at variance with the results obtained in an InSb QW having still larger $g^*m^*/m_0$ value, where  considerably sublinear trajectories of this kind were found.\cite{Nedniyom,Yang-Santos} That meant a decreased effective $g$-factor with decreased $B_{\bot}$ and a nonlinear dependence of spin polarization on magnetic field. Linear trajectories in the HgTe QW mean, conversely, a field independent $g^*m^*/m_0$ and a linear field dependence of the spin polarization. It is noteworthy that the found difference in spin-polarization properties of InSb and HgTe QW is not due to a small effective mass in InSb, as was suggested in Ref.\onlinecite{Yang-Santos}, since the corresponding mass here $m^*/m_0=0.024$ (Ref.~\onlinecite{Kvon}) is not radically different from that for the InSb QW $m^*/m_0=0.018\div0.026$.\cite{Nedniyom} This is a topic for future investigations to understand whether this difference is due to the $\Gamma_6$ and $\Gamma_8$ characters of conduction bands in these materials or due to a relatively wide QW of 30~nm in InSb\cite{Yang-Santos} when the magnetic length equals the well width already at 0.75~T.

\begin{figure}[t]
\includegraphics[width=\columnwidth]{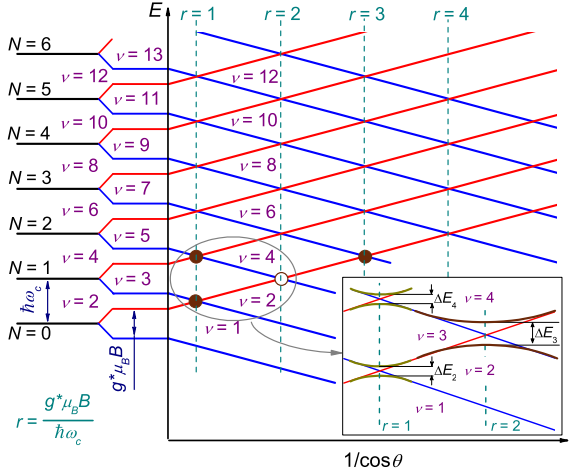}
\caption{\label{fig:levels}  (color online).  Evolution of the magnetic levels with tilt $\theta$ in a one-electron approach for a $\Gamma_6$-like conduction band. Filled circles correspond to the experimentally observed  stronger coincidence features while the open circle corresponds to the vanishing one. Inset: Anticrossings that substitute the level crossings in a one-electron picture as revealed experimentally (unscaled and exaggerated).}
\end{figure}

Second, another system of straight traces along the coincidences, descending from a single point $B_{||}^0$ on the $B_{||}$ axis, is well seen in the experimental pictures as a set of ridges going through the MR peaks: Figs. \ref{fig:XX-3D anticross}, \ref{fig:DescTraj}, \ref{fig:XX-XY}, and \ref{fig:Bpar}(a). The origins of the two sets of trajectories may be understood on the basis of the level coincidence scheme (Fig.~\ref{fig:levels}) drawn for a fixed $B_\bot$ as a function of tilt $1/\cos{\theta}$. Here each of the first-type trajectories corresponds to a set of level coincidences along a dashed vertical for a fixed $\theta =\theta_r$. The coincidences at even (odd) $\nu$  reside on the traces for odd (even) $r$ values. The second-type trajectories are formed by the sets of coincidences residing on any of the fixed upper spin sublevel (e.g., see the circles on the $N=0$ upper spin sublevel). Each group of such coincidences on the second-type trajectory is characterized by a fixed value of $M\equiv \nu-r= 1,3,5\dots$, and a corresponding trajectory on the $(B_{\bot},B_{||})$ plane is described in terms of the $\Gamma_6$-like band as follows:
\begin{eqnarray}
B_{||}=&&B_\bot \tan{\theta_r}=B_\bot \sin{\theta_r}\frac{2r}{g^*m^*/m_0}=\nonumber\\
&&B_\bot \sin{\theta_r}\frac{2(\nu-M)}{g^*m^*/m_0}=
B_\bot \sin{\theta_r}\frac{2(B_1/B_\bot-M)}{g^*m^*/m_0}=\nonumber\\
&&2\sin{\theta_r}\frac{(B_1-MB_\bot)}{g^*m^*/m_0}\approx2\frac{(B_1-MB_\bot)}{g^*m^*/m_0}
\label{eq:Bpar-Bper}
\end{eqnarray}
for high enough tilts such that $\sin{\theta_r}\approx1$. 
Here $B_1=hn_S /e$ is the $\rho_{xx}(B_\bot)$ minimum position for $\nu=1$. It is notable that a similar equation for the $\rho(B_{\bot},B)$ graph with the total field used instead of $B_{||}$, usually built in other works (e.g., in Ref.~\onlinecite{Yang-Santos}), holds exactly for the whole range of $\theta$. The descending straight beams in Fig.~\ref{fig:XX-XY} are built according to this equation and a good overall agreement with experimental data is seen. The curve declining from the $M=1$ trace is built without the $\sin{\theta_r}=1$ approximation, and still better agreement is obtained for it as it goes through the crossing of the $r=1$ ascending trace with the $\nu=2$ vertical. It is remarkable that coincidences for {\it all} successive filling factors reside on a descending trace while only odd- or even-numbered coincidences reside on a traditional ascending trace. That is why the second type traces are better seen in the experimental pictures since the neighboring coincidence peaks overlap along them forming quasicontinuous chains or ridges. 

\begin{figure}[h]
\includegraphics[width=\columnwidth]{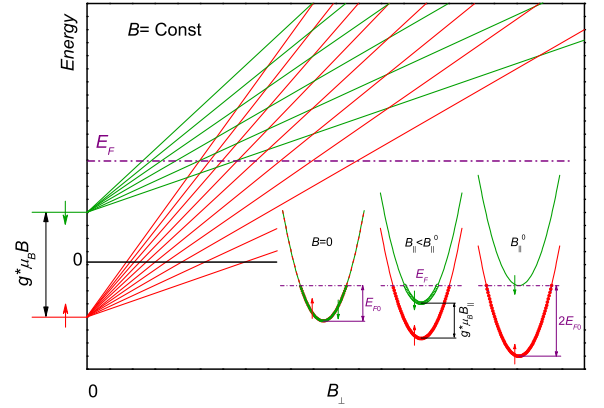}
\caption{\label{fig:Spin-polarization}  (color online). A schematic presentation of magnetic levels in a $\Gamma_6$-like band as a function of $B_\bot$ for a scan of the $(B_\bot,B_{||})$ plane along a circle trajectory with a fixed $B$. Inset: redistribution of electrons between two spin subbands under parallel fields resulted in the full spin polarization of the electronic system at $B_{||}\geq B_{||}^0$.}
\end{figure}

The expression for the convergence point $B_{||}^0=2B_1/(g^*m^*/m_0)$ in Eq.(\ref{eq:Bpar-Bper}) may be transformed into another form: $g^*\mu_{B}B_{||}^0= 2E_{F0}$ ($E_{F0}=\pi\hbar^2n_S/m^*$ denotes the Fermi level for a spin degenerate case, $\mu_B$ denotes the Bohr magneton), thus making clear the essence of this point---it is the field of a full spin polarization of the electronic system (Fig.~\ref{fig:Spin-polarization}, inset).

\begin{figure}[h]
\includegraphics[width=\columnwidth]{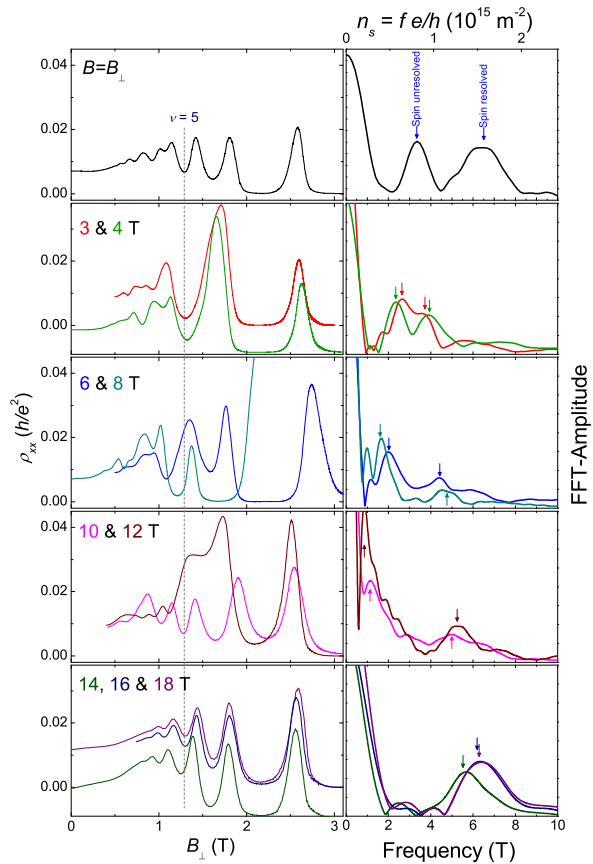}
\caption{\label{fig:Rxx-FFT-Sum}  (color online). $\rho_{xx}$ oscillations (left) taken from the  $(B_\bot,B_{||})$ map of Fig.~\ref{fig:XX-XY}(a) and their Fourier transforms (right). Upper plots are for oscillations under pure perpendicular fields and the others are taken along circle trajectories for increasing radius $B$. }
\end{figure}

\begin{figure}[t]
\includegraphics[width=\columnwidth]{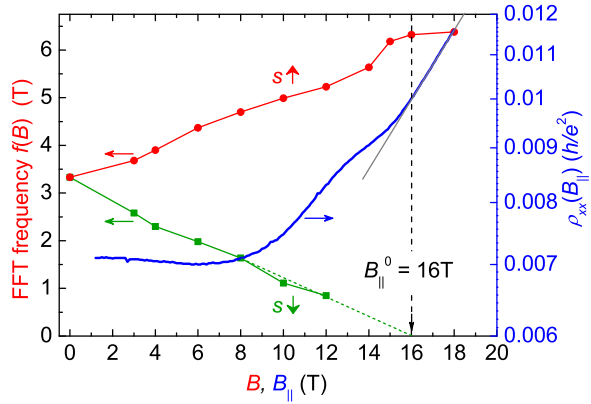}
\caption{\label{fig:FFT-Res}  (color online). Fourier frequencies along circle trajectories as functions of $B$ compared to the position of the convergence point $B_{||}^0$ (dashed vertical) of the descending trajectories and to MR under a pure parallel field.}
\end{figure}

\subsection{\label{sec:level5}Spin polarization as deduced from circle orbits in the $(B_\bot,B_{||})$ plane and from MR under pure parallel fields}

Another way to estimate the redistribution of electrons between two spin subbands is to perform a Fourier analysis [fast Fourier transform (FFT)] of oscillations in  $\rho_{xx}(1/B_\bot)$ taken along the circle trajectories in the $(B_{\bot},B_{||})$ plane for fixed values of the total field $B$ (Refs. \onlinecite{Fang-Stiles} and  \onlinecite{Tutuc-2002}) (Fig.~\ref{fig:Rxx-FFT-Sum}). In this case, only $B_{\bot}$ is a variable in the description of magnetic levels $E_{N,s}=(N+1/2)\hbar eB_{\bot}/m^*\pm g^*\mu_BB/2$ while $B$ is constant (Fig.~\ref{fig:Spin-polarization}). This yields two spin series periodic in $1/B_{\bot}$ for $E_{N,s}=E_F>g^*\mu_BB$ resulting in two lines of FFT frequencies $f_i$~{\it vs}.~$B$ (Fig.~\ref{fig:FFT-Res}), which describe electron densities in the two spin subbands as $n_{Si}=f_ie/h$. The lines diverge from a single point, which corresponds to a lower frequency peak in the FFT diagram for the pure perpendicular fields (Fig.~\ref{fig:Rxx-FFT-Sum}, upper right) that relates to the low field range where the spin peaks are unresolved, while the upper-frequency peak on this plot is for the higher field range with resolved spin splittings. The lower line in Fig.~\ref{fig:FFT-Res} goes to zero  meaning the exhausting of the upper subband, and the upper line simultaneously saturates at $f$ corresponding to $n_{s\uparrow}=n_{S}$, thus indicating the moment of the full spin polarization. 

The above results are compared in Fig.~\ref{fig:FFT-Res} with MR in a pure $B_{||}$ and it is seen that just the field of a full spin polarization separates two kinds of dependences in  $\rho_{xx}(B=B_{||})$: a somewhat complicated function at lower fields, where two spin subbands are filled, from a monotonously increasing one, for a single subband filled, at higher fields as has been observed in a number of studies.\cite{Nedniyom,Yang-Santos,Yoon,Drichko}

\subsection{\label{sec:level4}On the $g$-factor anisotropy}

Under pure perpendicular fields, the odd-numbered QH features in the investigated structure were found to prevail over the even-numbered ones (see Fig.~\ref{fig:Rxx-FFT-Sum}, upper left, and Ref.~\onlinecite{YaPE}) meaning that spin gaps are larger than cyclotron gaps. It is notable that such a result was found in Ref.~\onlinecite{Zhang-2004} only for structures with asymmetric one-sided doping, while for structures with symmetric two-sided doping, a traditional relation for spin gaps smaller than cyclotron gaps (odd-numbered features weaker than even-numbered ones) has been obtained. This contradicts our results, as our sample is symmetrically doped, and observation of straight trajectories along the coincidences in the $(B_\bot,B_{||})$ plane confirms it. The source of the discrepancy may be in that our sample has a larger mobility (22 versus $7.35\times10^4$~cm$^2$/Vs) at lower densities (1.5 versus $6.59\times10^{11}$~cm$^{-2}$) and wider QW (20.3 versus 9~nm). Then the spin gaps may be exchange enhanced\cite{Ando-Uemura} in our case reversing the relation between the spin and cyclotron gaps, while this effect is weaker and insufficient for the symmetric sample of Ref.~\onlinecite{Zhang-2004}. 

\begin{figure}[b]
\includegraphics[width=9cm]{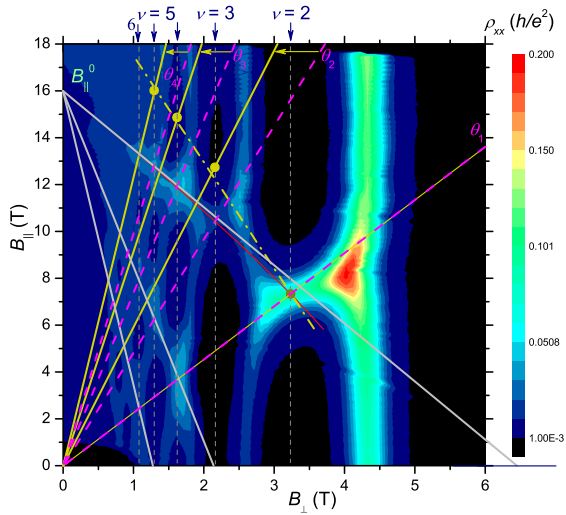}
\caption{\label{fig:XX_g-anisotr}  (color online). A picture of coincidences in a state as in Fig.~\ref{fig:XX-XY}(a) with added ascending traces (solid beams) for the estimated anisotropic $g$ factor. They are connected by the horizontal arrows with corresponding traces for the isotropic $g$ factor (dashed beams). The ascending traces for both cases are fitted to coincide for $r=1$, but the traces for anisotropic $g^*$ deviate to larger tilts for $r>1$. Correspondingly, the expected coincidence peak positions ($\bullet$) deviate considerably to larger $B_{||}$ from the experimental positions, and the descending trace through them (dash-dot line) should dramatically increase its slope.}
\end{figure}

In terms of the $\Gamma_6$ band, the prevalence of spin gaps should indicate the relation $g^*m^*/m_0>1$, and the value $g^*m^*/m_0=1.38$ was indeed found from the fields of oscillation onset and the onset of their splittings.\cite{YaPE} On the other hand, the condition $g^*m^*/m_0>1$ means that the first ascending trajectory should go at the angle $\theta_1<60^\circ$ while our experiment yields $\theta_1=66.2^\circ$, corresponding to $g^*m^*/m_0=0.807$, and this result is in good overall experimental agreement with other ascending trajectories for $r>1$ drawn in Figs.~\ref{fig:XX-XY} and \ref{fig:XX_g-anisotr} according to a condition 
\begin{equation}
\cos\theta_r=\cos\theta_1/r
\label{eq:asctrisrel1}
\end{equation}
that follows from Eq.~(\ref{eq:trad}).

This problem could hardly be resolved by consideration of the efective mass dependence on $B_{||}$\cite{Smr-J} since the $g$-factor is inversely proportional to $m^*/m_0$ for small masses thus the $g^*m^*/m_0$ product would be incensitive to the mass variation, at least in the first spproximation.  Initially,\cite{YaPE} we tentatively explained the discrepancy between the prevailing spin gaps under pure perpendicular fields and the value of $\theta_1>60^\circ$ in terms of the $g$-factor anisotropy.\cite{Ivchenko} It is introduced in the form:
\begin{equation}
g^{*2}(\theta)=g_z^2\cos^2{\theta}+g_{xy}^2\sin^2{\theta}
\label{eq:g-anisotropic}
\end{equation}
with $g_z$ and $g_{xy}$ being the $g$-factor components for the field orientations perpendicular and parallel to the layers, correspondingly. Such an explanation looked plausible as $g_{xy}$ is known to be much smaller than $g_{z}$, even approaching zero, for the $\Gamma_8$ valence band in traditional semiconductor QWs.\cite{Kesteren-Marie} Then, considering the oscillations under perpendicular fields and solely the coincidences on the $r=1$ trace with the corresponding value of $\theta_1$, a certain value of anisotropy was deduced. 

But now, after measurements up to higher fields and obtaining the distinct positions for ascending traces with $r>1$, we see that situation becomes more complicated if one takes into account the whole picture of coincidences. To describe the whole set of ascending traces for an anisotropic $g^*$-factor (\ref{eq:g-anisotropic}), the traditional Eq.~(\ref{eq:trad}) is modified to
\begin{equation}
\tan^2{\theta_r}=\frac{4r^2}{(g_{xy}m^*/m_0)^2}-\left( \frac{g_z}{g_{xy}} \right)^2,
\label{eq:asctraces-anisotropic}
\end{equation}
which transforms into (\ref{eq:trad}) for an isotropic $g^*$ with $g_z=g_{xy}$. Equation~(\ref{eq:asctraces-anisotropic}) yields a set of straight beams for $r=1,2,3...$ going from zero in the $(B_\bot,B_{||})$ plane, as for the isotropic $g$, but it is obvious that the beams for larger $r$ should decline from the corresponding traces for the isotropic $g$ to larger tilts, approaching the $B_{||}$ axis, as $g^*m^*/m_0$ will decrease with increased $r$. To describe the relative positions between ascending trajectories  for anisotropic $g^*$, the Eq.(\ref{eq:asctrisrel1}) is modified to:
\begin{equation}
\frac{1}{\cos^2{\theta_r}}=\frac{r^2}{\cos^2{\theta_1}}+\left[ \left( \frac{g_z}{g_{xy}} \right)^2-1 \right](r^2-1).
\label{eq:asctranisrel1}
\end{equation}
For the known values of $\theta_1$ and $g_zm^*/m_0=1.38$, the value of $g_{xy}m^*/m_0$ to fit the first ascending trajectory is estimated from Eq.~(\ref{eq:asctraces-anisotropic}) as
\begin{equation}
g_{xy}m^*/m_0=\frac{ \sqrt{ 4-(g_zm^*/m_0)^2 } } { \tan{\theta_1} }=0.638,
\label{eq:asctranis1}
\end{equation}
yielding the value of $g_z/g_{xy}=2.16$.\cite{Comment1}

In Fig.~\ref{fig:XX_g-anisotr}, with the same experimental data as in Fig.~\ref{fig:XX-XY}(a), two sets of ascending trajectories for the isotropic and anisotropic $g^*$ are compared. The coincidences should appear at the cross points of the calculated traces and the verticals for the integer filling factors. It is seen that, once there is a good overall agreement between experimental and calculated coincidences in the model of isotropic $g^*$, the agreement is impossible for the $g$-factor anisotropy of the deduced value, with disagreements being well beyond the possible experimental errors. Especially large is a deviation of the descending trajectory drawn through the expected coincidence points for anisotropic $g^*$, as its extrapolated crossing with the $B_{||}$ axis goes far beyond the $B_{||}^0$ point in the isotropic model, thus totally destroying the agreement with the field of full spin polarization found above by other means.

Thus the anisotropic $g^*$ in the form of Eq.~(\ref{eq:g-anisotropic}) does not explain the found discrepancy between oscillation under perpendicular fields and the data found from coincidences in tilted fields. If the large value of $g_{z}m^*/m_0>1$ under perpendicular fields is due to the exchange enhancement of spin gaps,\cite{Ando-Uemura} then either a small addition of $B_{||}$ quickly destroys the effect or the difference in physics for oscillations under pure perpendicular fields and at the coincidences is essential. For example, a spin gap in the former case is exchange-enhanced when the Fermi level is within it, between two spin sublevels of the same Landau level, while coincidences always occur for spin levels of {\it different} Landau levels. More detailed theoretical analysis with many-body effects included is needed to verify this.

\subsection{\label{sec:level6}Structures in MR close to parallel fields}

\begin{figure}[h]
\includegraphics[width=10.5cm]{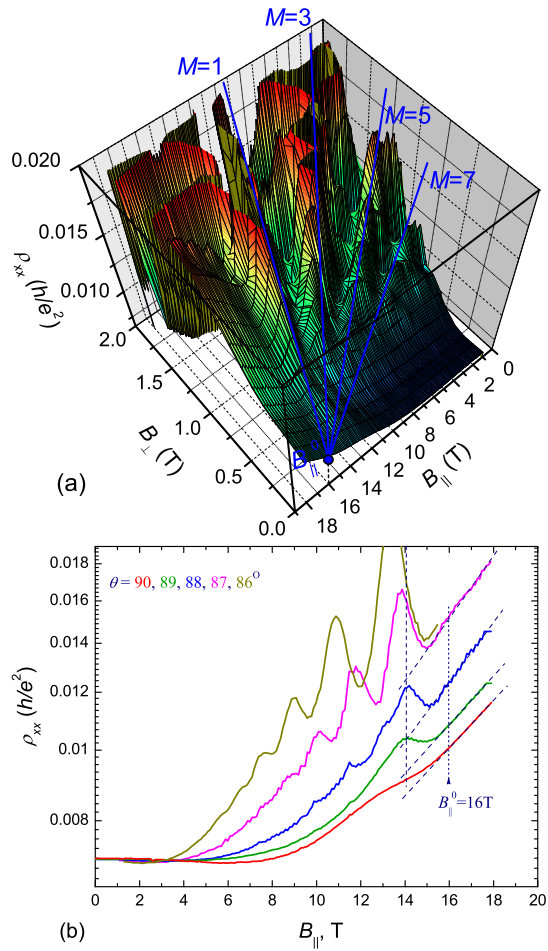}
\caption{\label{fig:Bpar}  (color online). (a) The same as in Fig.~\ref{fig:XX-3D anticross}(b) but at small $B_{\bot}$ and viewed from $B_{||}^0$. (b) Cross-sections of (a) taken at $\theta=90^{\circ}(B_{||}),89,88,87,86^{\circ}$.}
\end{figure}

The structure seen in $\rho_{xx}(B=B_{||})$ at $B<B_{||}^0$ (Fig.~\ref{fig:FFT-Res}) is enhanced with small deviations of the field from the parallel orientation [Fig.~\ref{fig:Bpar}(b)], as was observed in InSb QWs.\cite{Nedniyom,Yang-Santos} Contrary to the results in InSb, we cannot state that we see a single peak under quasiparallel fields. Rather, we see an emerging series of coincidence peaks positioned on the second-type trajectories in the $(B_{\bot},B_{||})$ plane for $M=1,3,5...$. On the other hand, it is senseless to speak of coincidences under strictly parallel field, thus a wide maximum at $B\approx14$~T should be of different nature. Its position is insensitive to small tilts of the field from the parallel orientation, in disagreement with some shift to lower $B_{||}$ of the $M=1$ trajectory (Fig.~\ref{fig:XX-XY}), and only at the field deviations larger than  $3^\circ$ do we see the peaks start moving to lower $B_{||}$. Thus an interference of some processes causing MR under strictly parallel fields (e.g., scattering between spin subbands\cite{Nedniyom,Yang-Santos}) with the emerging coincidence features is observed in HgTe QW. Comparing the results in quasi-parallel fields for InSb and HgTe QWs, the following conclusions may be drawn: either the coincidence peaks are suppressed in InSb or the causes for a single peak in HgTe are weakened. The found strongly nonlinear coincidence trajectories in the InSb QW may promote the absence of the coincidence peak series, as the peaks would tend to converge at high-$B_{||}$ area in that case.

\subsection{\label{sec:level7}Anticrossings}

\begin{figure}[b]
\includegraphics[width=9.3cm]{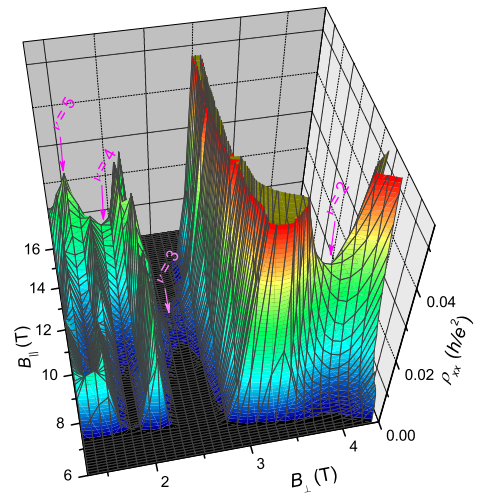}
\caption{\label{fig:IL1-CS1(nu3)cropped}  (color online). Enlarged part of Fig.~\ref{fig:XX-3D anticross}(c) demonstrating the vanishingly small coincidence feature for $\nu=3$ and $r=2$. Arrows indicate positions of coincidences along the $M=1$ descending trajectory.}
\end{figure}

\begin{figure}[t]
\includegraphics[width=\columnwidth]{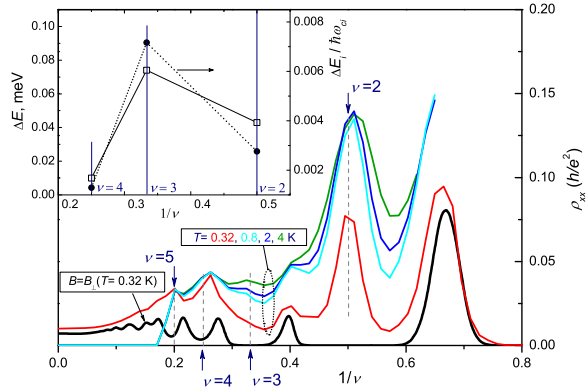}
\caption{\label{fig:M1crossect}  (color online). Cross sections of $\rho_{xx}(B_{\bot},B_{||})$ in the same sample state as in Fig.~\ref{fig:XX-XY}(a) along the exact $M=1$ trajectory at $T=0.32, 0.8, 2,$ and 4~K as compared to oscillations under pure perpendicular fields. Inset: The deduced activation energies at $\nu=2, 3,$ and 4 and their ratios to corresponding cyclotron energies.}
\end{figure}

At higher $B_{\bot}$, where the QHE is well developed, the MR features for coincidences acquire a complicated structure with local  $\rho_{xx}$ peaks splitted in couples of peaks shifted in opposite directions of $B_{\bot}$  onto the neighboring MR ridges and leaving decreased MR bridges at the points of expected level crossings [Figures \ref{fig:XX-3D anticross}, \ref{fig:DescTraj}, and \ref{fig:XX-XY}(a)], which are in fact "the spikes"\cite{Spikes} in the $\rho_{xx}(B_{\bot},B_{||})$ cross-sections for fixed $\theta$. This indicates the formation of anticrossings at the points of expected level coincidences (Fig.~\ref{fig:levels}). The effect is significantly enhanced after ir illumination due to a considerable improvement of oscillations (Fig.~\ref{fig:XX-3D anticross}). Unexpectedly, it was found that the anticrossings depend nonmonotonously on $B_{\bot}$: the anticrossing at $\nu=3$ is pronouncedly stronger than the neighboring ones at $\nu=2$ and 4, positioned along the same descending trajectory for $M=1$ [Figs. \ref{fig:XX-3D anticross}(b), \ref{fig:XX-3D anticross}(c), \ref{fig:DescTraj}, \ref{fig:XX-XY}, \ref{fig:IL1-CS1(nu3)cropped}, and \ref{fig:M1crossect}]. This observation illustrates another  useful property of the newly found descending trajectories, as the coincidence features for {\it all} successive filling factors should exist on each of them within the one-electron model, as opposed to those along the traditional ascending trajectories, where the $\nu=3$ coincidence fundamentally should not exist on the $r=1$ trace but only on the $r=2$ trace. The $\nu=3$ anticrossing is dramatically enhanced after illumination (Fig.~\ref{fig:XX-3D anticross}). This indicates a high sensitivity of the anticrossings to the changes in the sample quality, which is typical for electronic phase transitions.\cite{Jungwirth-McD,Koch-Desrat} The activation gaps deduced from the temperature dependences of MR at anticrossings (Fig.~\ref{fig:M1crossect}) confirm the nonmonotonicity, with the $\nu=3$ gap being more than half an order of magnitude larger than those for its neighbors. This result looks counterintuitive since the overlapping of magnetic levels with decreased $B_{\bot}$ seems to monotonously destroy the causes for the appearance of anticrossings, suppressing the spin polarization, as has been observed so far on other materials.\cite{Koch-Desrat,Jungwirth98}

A conventional explanation of the anticrossings is in terms of the magnetic anisotropy in the electronic system\cite{Jungwirth-McD,Jungwirth98}: as the approaching spin levels of different spin orientations swop their order in energy relative to the Fermi level, the Hartree-Fock energy of the system may decrease due to changes in the many-body interaction concomitant to changes in the spin polarization. The decrease starts before the level crossing, due to a hybridization of the levels, and the crossing in fact does not occur. The stronger the energy gain due to spin polarization is, the larger is the anticrossing gap. The systems reacting to transitions into spin ordered states under QHE conditions are the QH Ferromagnets (QHF). Estimations for an easy-axis QHF in a 2D layer\cite{Jungwirth-McD} do not yield a sensible difference for anticrossings at $\nu= 2, 3$, and 4. That is why we tentatively attribute the observed difference to the coupling of $B_{||}$ with the orbital degree of freedom in a QW of a finite width resulted in a different charge density profile across the QW for the two approaching spin levels.\cite{Jungwirth98} This coupling enhances a magnetic anisotropy energy for the $\nu=3$ coincidence as compared to that for $\nu= 2$ since its $B_{||}$ is about a factor of 1.5 higher, resulting in a shrinkage of the wave functions and thus the increased spatial difference between them. On the other hand, the coincidences at $\nu\geq4$ are restored because they go outside of the QH range of $B_{\bot}$.\cite{Koch-Desrat,Jungwirth98}

\section{Conclusions}
In summary, we have found that the coincidence features in MR of a symmetrical HgTe QW under tilted fields form a set of descending straight trajectories in the $(B_{\bot},B_{||})$ plane diverging from a single point that yields a field of the full spin polarization of the electronic system, which is in agreement with the values obtained in other ways. The $g$ factor found under pure perpendicular fields is twice as large as  the value found from coincidences, and we show that this difference cannot be described  in terms of a conventional $g$-factor anisotropy since the whole  picture of coincidences is well described with the isotropic $g$ factor while an introduction of its anisotropy inevitably destroys this agreement. The anticrossings of spin levels that appear in the QH range of fields were found to behave dramatically nonmonotonously with field, demonstrating the intricate nature of the emerging QHF phase. 

\begin{acknowledgments}
Authors are grateful to E. Palm, T. Murphy, J. H. Park, and G. Jones for help with the experiment. Supported by RFBR, project 11-02-00427, and by the Program of UD RAS, project No 12-P-1-100. National High Magnetic Field Laboratory is supported by NSF Cooperative Agreement No.~DMR-0654118, the State of Florida, and the US DOE.\end{acknowledgments}


\begin{thebibliography}{99}
\bibitem{Comment1} When estimating the $g$-factor anisotropy in Ref.~\onlinecite{YaPE} we used another relation $g=g'_\bot\cos\theta+g'_{||}\sin\theta$ and obtained the value: $g'_\bot/g'_{||}\simeq 5$ for the same case, but this is in fact a different parameter.
\end{thebibliography}
\end{document}